\documentclass[12pt]{article}
\input epsf.sty
\usepackage{amssymb}
\usepackage{amsmath}
\usepackage{comment}
\usepackage{mathrsfs}
\usepackage[all]{xy}
\usepackage[dvips]{graphicx}
\usepackage{epsfig,color}
\usepackage{amsmath}
\usepackage{physics}
\usepackage{setspace}
\usepackage{fancybox}

\def\be{\begin{equation}}
\def\ee{\end{equation}}
\def\ba{\begin{align}}
\def\ea{\end{align}}
\def\a{\alpha}
\def\b{\beta}
\def\p{\partial}

\def\ops[#1]{\p_{#1} e^{-2\phi}}

\def\GT{\rightarrow}

\def\eq[#1]{equation (\ref {eq:#1})}
\def\Eq[#1]{Equation (\ref {eq:#1})}
\def\e[#1]{\ref {eq:#1}}
\def\at[#1]{| _{#1}}

\let\oldpercent\%\renewcommand{\%}{\scalebox{0.85}{\oldpercent}}

\begin{document}

\baselineskip=18pt

\begin{center}
{\Large \bf{ Instant Folded Strings and Black Fivebranes}}

\vspace{10mm}

\textit{ Amit Giveon$\,^1$, Nissan Itzhaki$\,^2$ and Uri Peleg$\,^2$}
\break

$^1$ Racah Institute of Physics, The Hebrew University \\
Jerusalem, 91904, Israel

$^2$ Physics Department, Tel-Aviv University, Israel	\\
Ramat-Aviv, 69978, Israel \\

\end{center}


\vspace{10mm}

\begin{abstract}

We study the recent claim that black NS5-branes are filled with folded strings. We calculate, in the near-extremal case,
the number of folded strings at the black fivebranes interior, using different approaches,
and get the exact same answer.
The backreaction of the folded strings leads us to argue that the interior of the black fivebrane is $AdS_2$ (times a compact manifold)
and that infalling matter cannot reach the interior, due to a shock wave at the horizon.
These considerations also suggest a novel insight into the black fivebranes entropy.
\end{abstract}
\vspace{10mm}

Recently, it was argued that black NS5-branes are not empty, but are filled with folded strings \cite{Itzhaki:2018glf}.
Unlike standard folded strings (see e.g. \cite{Bardeen:1975gx,Bardeen:1976yt,Bars:1994qm,Gubser:2002tv}),
those discussed recently are classical solutions that describe a closed folded string, which is created at an instant.
We, therefore, refer to them as Instant Folded Strings (IFS).
The fact that they are created classically at an instant, leads to several unusual features.
Perhaps the most notable is the fact that they violate the Averaged Null Energy Condition (ANEC) \cite{Karinne}.
This suggests that their backreaction on the black fivebranes geometry is rather non-trivial,
and that it might lead to interesting insights about black-hole physics.

To explore this, we have to know the number of IFS's, $N_{IFS}$. There are two reasons to expect this number to scale like $1/g_s^2$,
where $g_s$ is the string coupling near the horizon, which we take to be parametrically small.
The first follows from the  exact coset CFT description of the near-horizon limit~\cite{Maldacena:1997cg} of $k$ black NS5-branes --
the $SL(2,\mathbb{R})_k/U(1)$ black-hole model~\cite{Bars:1990rb,Witten:1991yr,Dijkgraaf:1991ba,Elitzur:1991cb,Mandal:1991tz}
(times a compact space).
Using \cite{Giribet:2000fy,Giribet:2001ft}, the operator that corresponds to the folded strings was identified in \cite{Itzhaki:2019cgg,Giveon:2019gfk}.
It was shown that this operator receives an expectation value on the sphere,
and since its wave function is highly localized near the horizon,
it implies that $N_{IFS}\sim 1/g_s^2$.
A more direct argument is that the topology associated with the creation of the IFS is that of a disc. Hence, the amplitude for an IFS creation scales like $1/g_s$, and the production rate like $1/g_s^2$.

The fact that $N_{IFS}\sim 1/g_s^2~$ implies that their backreaction should modify the interior of the black NS5-brane geometry considerably.
To determine the IFS effects, we have to know the exact prefactor.
A direct calculation of the prefactor  is beyond the scope of this note.
Instead, here we present different {\it indirect} approaches to calculate $N_{IFS}$,
and find the exact same answer,
\be\label{number}
N_{IFS}=\frac{2\pi }{k g_s^2}~.
\ee
In the process, we expose the fascinating role that the IFS play in the black fivebranes dynamics.

The setup we consider is the familiar
near-horizon limit of black $k$ NS5-branes \cite{Maldacena:1997cg}. It is described by the background
\be
SL(2,\mathbb{R})_k/U(1) \times  S^3 \times T^5.
\ee
The $S^3$ and $ T^5$ parts of the background play no role here.
Moreover, for the most part, the exact $SL(2,\mathbb{R})_k/U(1)$ CFT description is not necessary for the considerations presented next.

It is sufficient for our purposes to inspect the following two-dimensional space-time effective action:
\be S=\frac{1}{2}\int d^2x \sqrt{-g} \ e^{-2\phi} [R + 4(\nabla \phi)^2 + 4Q^2] \ +S_{IFS}~. \label{eq:action} \ee
The first term is the two-dimensional dilaton-gravity action, with~\footnote{Here, we define $Q$ such that it has mass dimension $1$.}
\be
Q^2=\frac{1}{\alpha' k}~,
\ee
and the second term is the source term associated with the IFS, that will be described momentarily.
The resulting equations of motion are:
$$  4\nabla^2 \phi - 4(\nabla \phi)^2 + R + 4Q^2=0~, \label{eq:EOM1} $$
\be R^{\mu\nu} + 2\nabla^{\mu} \nabla^{\nu} \phi  = e^{2\phi} \ T_{IFS}^{\mu\nu}~,
\label{eq:EOM2} \ee
where $T_{IFS}^{\mu\nu}$ is the energy-momentum tensor associated with the IFS.

To study these equations, we follow \cite{Callan:1992rs}, and use  the gauge
\be
g_{+-}=-\frac12 \exp(2\rho),~~~g_{--}=g_{++}=0~.
\ee
The equations of motion in the absence of IFS then take the form
$$4\p_{+}\rho\p_{+}\phi-2\p_{+}^2\phi=0 \ , \ \ \ 4\p_{-}\rho\p_{-}\phi-2\p_{-}^2\phi=0~,$$
\be \p_{-} \p_{+} e^{-2\phi} -  e^{2\rho-2\phi}Q^2 =0 \ , \ \ \  \p_-\p_+(\rho-\phi)=0~. \label{eq:vacEOM}\ee
The two-dimensional black hole is the solution
\be\label{bhs} e^{-2\rho}=e^{-2\phi}= \frac{M}{Q}-Q^2x^-x^+~,\ee
where $M$ is the ADM mass of the black hole. The horizon is at $x^{\pm}=0$,
and the singularity at the hyperbola $x^-x^+=M/Q^3$. The entropy of the black hole is
\be\label{sbh}
S_{BH}=2\pi e^{-2\phi_0}\equiv{2\pi\over g_s^2}~,
\ee
where $\phi_0$ is the value of the dilaton at the horizon (see e.g. \cite{Giveon:2005mi}, for a review).

Locally behind the horizon, the solution takes the form~\footnote{Up to corrections in $1/k$; see~\cite{Karinne} for details.}
\be\label{bac}
ds^2=-(dX^0)^2+(dX^1)^2,~~~\Phi=\Phi_0+\tilde{Q} X^0,
\ee
with $Q>\tilde{Q}>0$.~\footnote{The exact expression for $\tilde{Q}$ can be found in \cite{Karinne}, and is not important here.}
Namely, the dilaton gradient is timelike, and it points towards the future.

In such a situation, IFS's are created \cite{Itzhaki:2018glf}. This follows from a trivial analytic continuation of \cite{Maldacena:2005hi}, which implies that locally (up to length scales set by the curvature -- $\sqrt{k\alpha'}$ in our case), the solution associated with an IFS that is created at $X^0=x^0$ and $ X^1=x^1$ reads
\be
X^1=\sigma; ~~~ X^0=x^0+\a'\tilde{Q} \log(\frac{1}{2}\left[ \cosh(\frac{\sigma-x^1}{\a'\tilde{Q}})+\cosh(\frac{\tau}{\a'\tilde{Q}})\right] ).
\ee
We are interested mainly in the large $k$ limit, which corresponds to $\tilde{Q}\to 0^+$.
In that limit, the energy-momentum tensor associated with an IFS that is created at $u=v=0$ is particularly simple,
\be
T_{uu}=-\frac{v}{2\pi\a'}\Theta(v)\delta(u)~, ~~~T_{uv}=\frac{1}{2\pi\a'}\Theta(v)\Theta(u)~,~~~T_{vv}=-\frac{u}{2\pi\a'}\Theta(u)\delta(v)~,
\label{eq:T}
\ee
where $u=x^0-x^1$ and $v=x^0+x^1$.

At first, this energy-momentum tensor looks  standard: it is conserved, away from the edges of the IFS only $T_{uv}$ does not vanish, and it describes the energy-momentum tensor  of a folded string (with twice the tension). However, the total energy (and momentum) associated with it vanish. This is not by accident. Before $t=0$, there was no string, so the energy and momentum were zero. By energy-momentum conservation, the string that was created at $t=0$ must have vanishing energy and momentum.

The fact that the total energy vanishes implies that $T_{uu}$ and $T_{vv}$ must be negative and that the ANEC is violated.
Moreover, since as we increase $t$ the contribution of $T_{uv}$ to the total energy grows, the amount of negative null flux at the edges must increase as well. This plays a crucial role below.

Now, we are ready to determine $N_{IFS}$. The first indirect way to do so is the following.
The cause for the IFS creation is the local timelike dilaton. Since the rate of production of the IFS is so large, $\sim 1/g_s^2$,
it is natural to suspect that $N_{IFS}$ is determined by asking how many IFS's are needed to render the dilaton constant.
For this, it is sufficient to look at the bulk of the IFS away from the edges,
where the only non-vanishing component of the energy-momentum tensor is $T_{uv}$.
Plugging (\ref{eq:T}) into the right hand side of (\ref{eq:vacEOM}), we see that  the two non-trivial equations are
$$ \p_{-} \p_{+} e^{-2\phi} + e^{2\rho-2\phi}Q^2 = T_{-+} =  \frac{e^{2\rho}N_{IFS}}{2\pi\a'}~, $$
\be
2e^{-2\phi}\p_-\p_+(\rho-\phi)=-T_{-+} = -\frac{e^{2\rho}N_{IFS}}{2\pi\a'}~. \label{eq:critical} \ee
The extra factor of $e^{2\rho}$ on the right hand side  follows from the non-trivial background.
The first equation implies  that to render the dilaton constant, $N_{IFS}$ must be eq. (\ref{number}).
Note that the other two equations that appear in (\ref{eq:vacEOM}) are satisfied  when the  dilaton is constant.

From the second equation above, we get that
\be
e^{-2\rho}=\frac{1}{2}Q^2(u-v)^2~.\ee
This implies that the metric is
\be ds^2 =\frac{2}{Q^2r^2}(-dt^2+dr^2)~,
\ee
which interestingly enough  describes an $AdS_2$ background (with a constant string coupling).
This suggests that the two-dimensional black-hole interior is replaced by $AdS_2$.
Needless to say that it would be nice to know if, in the sprit of  \cite{Maldacena:1997re}, there is a  QM dual to string theory on this non-supersymmetric $AdS_2$.

A different way to determine $N_{IFS}$ relies on the following reasoning \cite{Karinne}. The IFS violates the ANEC. In flat space-time,
this implies violation of unitarity \cite{Faulkner:2016mzt} and causality \cite{Hartman:2016lgu}.
Since the IFS are created behind the horizon, that would mean that unitarity and causality are violated inside the black hole.
A  way to avoid this conclusion is that the backreaction of the IFS is so strong, that they prevent particles from falling into the region
where unitarity and causality are violated. That is, since their rate of production is so large, $\sim 1/g_s^2$, they are created just inside the horizon and  cloak the black NS5-branes.

The energy-momentum tensor of the IFS has the right features to make this happen.
Suppose that $N_{IFS}$ IFS's are created just inside the black hole. Locally, the metric is flat and (\ref{eq:T}) can be used.
Since there is a negative null flux at the edges of the IFS, it generates a shock wave that pushes geodesics back in time,
\be
\Delta v \sim N_{IFS} \int du T_{uu}~.
\ee
Equation (\ref{eq:T}) implies that $\int du T_{uu}=-v$. Hence, one can fix $N_{IFS}$ by demanding that
\be\label{clo}
\Delta v= - v~.
\ee
In the appendix, it is shown that  (\ref{clo}) leads too exactly to (\ref{number}).

It is useful to understand why (\ref{clo}) implies that particles cannot enter the black hole.
For null geodesics, this is clear. A massless particle that attempts to cross the horizon, $u=0$, at some $v$, will be pushed,
according to (\ref{clo}), to $u=v=0$, and continue along the null geodesics $v=0$.

Timelike geodesics are  a bit more interesting. A massive particle that attempts to cross the horizon, $u=0$, at $v$,
will too be pushed to $u=v=0$. Naively, it will continue to follow its original trajectory into the black hole
(the dashed line in figure~1). However, (\ref{clo}) implies that it will be boosted too.
To see why, lets assume that $N_{IFS}$ is smaller than (\ref{number}). In that case,
$\Delta v=\alpha v$ with $\alpha=N_{IFS}k g_s^2/2\pi<1$. Then, the particle will not be pushed all the way to $u=v=0$, but to $u=0$ and $v=(1-\alpha)v$. This implies that the  wave length in the $v$ direction will also get smaller by a factor $1-\alpha$,
which implies that the particle is boosted by a factor $1/(1-\alpha)$ and that it follows the red line in figure~1.
For $\alpha=1$, this boost is maximal, and the particle will follow a null geodesics, $v=0$.

\begin{figure}
\begin{center}
\includegraphics[scale=0.45]{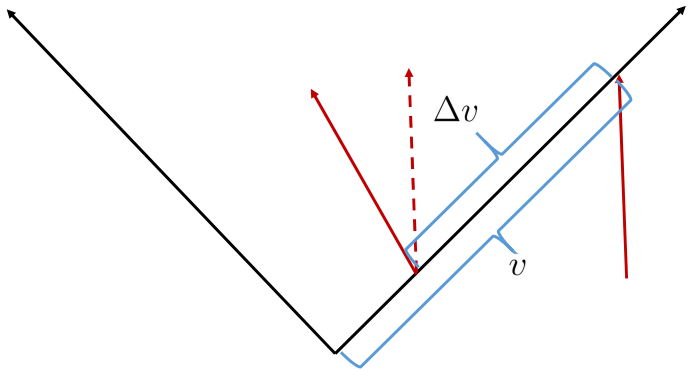}\vspace{-30mm}
\caption{A massive particle attempts to enter the horizon at some $v>0$ and $u=0$. The shock wave at the edge of the IFS at $u=0$ pushes the particle to $v+\Delta v$. The particle is also boosted by the shock wave, and it thus follows the solid line and not the dashed line. When (\ref{number}) is satisfied, $\Delta v=-v$, the boost factor is infinite, and the particle follows the $v=0$ trajectory -- it does not fall into the black hole. }
\label{SBHpen}
\end{center}
\end{figure}\label{fh1}

Thus, the black-hole interior, that was just argued to be $AdS_2$ space-time, is disconnected from the region outside the black hole,
due to a singular horizon (see figure~2), that prevents matter from falling in.
In~\cite{Ben-Israel:2017zyi,Itzhaki:2018rld}, a different reasoning led  to the same conclusion,
that the black NS5-branes horizon is singular.

Equation (\ref{number}) seems to also suggest a novel way to think about the black fivebranes entropy.
If black fivebranes are made of IFS's, then it is natural to suspect
that 
$S_{BH}\sim N_{IFS}$.
Very recently, \cite{Itzhaki:2019cgg,Giveon:2019gfk}, it was argued
that each IFS is 
a bound state of  $k$ gravitons (in a sense that is not entirely clear to us), which implies that
the total number of gravitons is
\be N_{gravitons}=kN_{IFS}~.\ee
This resonates with  \cite{Dvali:2011aa,Dvali:2012rt},
where it was suggested that the black hole should be viewed as a condensate of gravitons and
that  $S_{BH}\sim N_{gravitons}$.
Quite interestingly, the critical number of IFS's, found above and given in eq. (\ref{number}),
leads to an equality,
\be
N_{gravitons}={S_{BH}}~,
\ee
where the black-hole entropy is given in eq. (\ref{sbh}).

We view this as a further indication that perturbative string theory,
on the geometry of the exact $SL(2,\mathbb{R})_k/U(1)$ quotient CFT,
gives a concrete realization of
the idea that one can view classical black-hole geometries as quantum condensates
with large occupation numbers of soft gravitons \cite{Dvali:2011aa,Dvali:2012rt}.
Moreover, it points to the possibility that the gravitons condensates here are related to the gas of little strings
that dominate the black-hole thermodynamics at sufficiently high energy density
(see e.g.~\cite{Aharony:2004xn}, for a review),
and which appear in the microstates entropy counting done in~\cite{Maldacena:1996ya} and more recently in e.g.~\cite{Martinec:2019wzw}, in terms of strings living on the fivebrane.

\begin{figure}
\begin{center}
\includegraphics[scale=0.45]{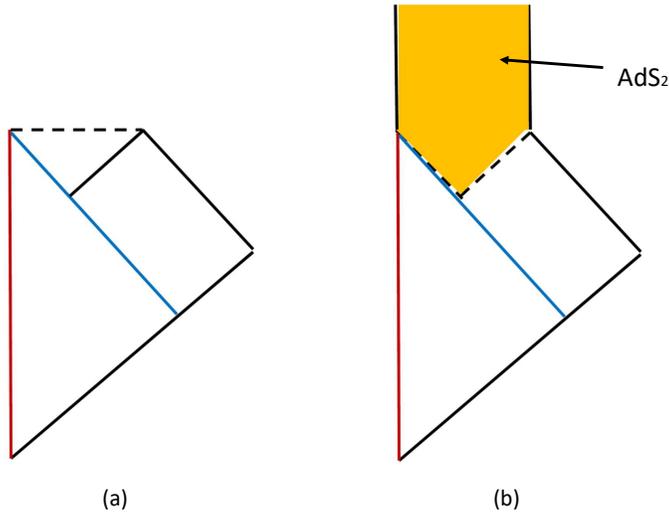}\vspace{-10mm}
\caption{
The Penrose diagram of a black hole formed by a null wave. (a) The standard picture: the blue line represents the null wave, the dashed line is the black-hole singularity and the red line is the strong string-coupling region. (b) The IFS picture: the region behind the horizon is filled with IFS's; asymptotically, it becomes $AdS_2$ space-time, and the horizon becomes singular in a way that prevents matter from falling into the $AdS_2$ region.}
\label{SBbHpen}
\end{center}
\end{figure}\label{fh1}

It is likely that the black NS5-branes entropy can be calculated also using  the IFS, without referring to gravitons. The IFS's are created behind the horizon, but not exactly at the same locations. It is reasonable to suspect that the black-hole entropy can also be associated with the $\log$ of the number of orthogonal wave functions of the IFS's. Moreover, it is reasonable to suspect that these wave functions are related to the non-trivial structure just behind the horizon of the black NS5-branes, discussed in~\cite{Ben-Israel:2017zyi,Itzhaki:2018rld}.
For this, however, a more detailed understanding of the IFS is required.

We would like to end with a comment about a limit that might be interesting to explore. The limit is
\be\label{limit}
 g_s\to 0,~~~~ k\to\infty, ~~~~ g_s^2 k\equiv\lambda~,
\ee
with finite $\lambda$. Naively, this limit seems to be trivial, as both the string coupling and curvature go to zero.
Indeed, this is the case for an external observer. For example, the only contribution to the reflection coefficient
of asymptotic states scattered from the black hole comes from the sphere in this limit. This contribution was calculated exactly in \cite{Teschner:1999ug} and, indeed, the stringy contribution to the reflection coefficient is trivial when $k\to \infty$.~\footnote{For
energies that scale with $\sqrt{k}/l_s$, the stringy effect on the scattering phase is dramatic, instead~\cite{Giveon:2015cma,Ben-Israel:2015mda}.}
This fits neatly also with the fact that the tail of the wave function of the IFS
decays exponentially with $\sqrt{k}/l_s$, away from the black hole (see~\cite{Itzhaki:2019cgg,Giveon:2019gfk} for details).

However, $N_{IFS}$ is finite in this limit, and it leads to the same effects:
it prevents infalling matter from crossing the horizon,
and the black fivebranes interior is replaced by an $AdS_2$ space-time.
This suggests that the limit (\ref{limit}) of string theory on the $SL(2,\mathbb{R})_k/U(1)$ black hole
can be used to focus on the IFS physics.

\vspace{10mm}

\section*{Acknowledgments}
This work is supported in part by a center of excellence supported by the Israel Science
Foundation (grant number 2289/18) and BSF (grant number 2018068).

\vspace{10mm}

\begin{appendix}

\section{ Shock wave at the edge of the IFS}
In this appendix, we calculate the shock wave at the edge of the IFS, and show how (\ref{number}) is obtained from (\ref{clo}).

We consider the backreaction of a single IFS on the two-dimensional black-hole solution (\ref{bhs}).
The IFS is created at the horizon bifurcation, $x^- =x^+ =0$.
The field equations are
$$e^{-2\phi}(4\p_{+}\rho\p_{+}\phi-2\p_{+}^2\phi)=-T_{++} \ , \ \ \ e^{-2\phi}(4\p_{-}\rho\p_{-}\phi-2\p_{-}^2\phi)=-T_{--}~,$$
\be \p_{-} \p_{+} e^{-2\phi} +  e^{2\rho-2\phi}Q^2 =T_{-+} \ , \ \ \ 2e^{-2\phi}\p_-\p_+(\rho-\phi)=-T_{-+}~.\ee
It is useful to parametrize the first order corrections to the background by
\be e^{-2\phi}\equiv \frac{M}{Q}-Q^2x^-x^+ + h~,\qquad \rho-\phi \equiv \omega~, \ee
where both $\omega$ and $h$ are  small.
The perturbative equations take the form
$$\p^2_+ h + 2Q^2x^-\p_+\omega=-T_{++}\ , \ \ \ \p^2_- h + 2Q^2x^+\p_-\omega=-T_{--}~,$$
\be \p_{-} \p_{+} h +  2Q^2\omega = T_{-+}\ , \ \ \ \ 2(M/Q-Q^2x^-x^+)\p_-\p_+\omega=-T_{-+}~.  \label{eq:EOMs} \ee
From the first three equations above, we get the following conditions on $h$:
\be (1-x^-\p_-)(\p_+^2 h) = -T_{++} - x^-\p_+ T_{-+}, ~~~ (1-x^+\p_+)(\p_-^2 h) = -T_{--} - x^+\p_- T_{-+}~. \label{eq:hdiff} \ee
The energy-momentum tensor we use is eq. (\ref{eq:T}), which after the appropriate coordinate transformation to the $\omega \equiv \rho-\phi=0$ gauge, gives
\be
T_{++} \cong -\frac{Qx^-}{2\pi\a'M}\Theta(x^-)\delta(x^+), ~~~T_{+-} \cong \frac{Q}{2\pi\a'M}\Theta(x^-)\Theta(x^+),~~~T_{--} \cong -\frac{Qx^+}{2\pi\a'}\Theta(x^+)\delta(x^-).
\ee
Here, we use ``$\cong$'' to denote equality up to leading order in $Q^2x^-x^+$.
Namely, eq. (\ref{eq:T}) is valid for distances of order $\sqrt{k}l_s$ from the horizon bifurcation.

\Eq[hdiff] now reads
\be (1-x^-\p_-)(\p_+^2 h) = 0, ~~~ (1-x^+\p_+)(\p_-^2 h) = 0~.\ee
The general solution is
\be h= \alpha x^+x^- +\beta x^- + \gamma x^+ +\delta ~, \label{eq:genh} \ee
for arbitrary coefficients $\alpha,\beta,\gamma,\delta $.
Near the horizon, we can approximate the last equation in (\e[EOMs]) as $$ \frac{2M}{Q}\p_-\p_+\omega \cong -T_{-+} \cong -\frac{Q}{2\pi\a'M}\Theta(x^-)\Theta(x^+)~,$$ and then integrate to solve for $\omega$
(while imposing the boundary condition for $\omega$ to be continuous),
\be \omega \cong -\frac{Q^2x^-x^+}{4\pi\a'M^2}\Theta(x^-)\Theta(x^+)~.  
\ee
We still have one more equation
(the third equation in (\ref{eq:EOMs})),
$$ \p_{-} \p_{+} h  = T_{-+} -  2Q^2\omega \cong \frac{Q}{2\pi\a'M}\Theta(x^-)\Theta(x^+)~,$$
that we can integrate to solve for $h$ (while imposing continuity of $h$)
\be h \cong \frac{Qx^-x^+}{2\pi\a'M}\Theta(x^-)\Theta(x^+)~.\ee
This is consistent with eq. (\ref{eq:genh}).
We now have all we need to determine the leading order corrections to the dilaton $e^{-2\phi}\cong \frac{M}{Q} - Q^2x^-x^+ +h $
and metric, $e^{-2\rho}\cong e^{-2\phi}(1-2\omega)$:
$$ e^{-2\phi}\cong
\frac{M}{Q}-Q^2x^-x^+ +\frac{Qx^-x^+}{2\pi\a'M} \Theta(x^-)\Theta(x^+)~,
\label{eq:phi} $$
\be e^{-2\rho} \cong \frac{M}{Q}-Q^2x^-x^+ +\frac{Qx^-x^+}{\pi\a'M} \Theta(x^-)\Theta(x^+)~.\label{eq:metric} \ee
The first equation describes the backreaction on the dilaton, that was discussed in the main text. Here, we focus on the second equation.

We are after the effects that take place when crossing the horizon $x^-=0$ at some finite $x^+>0$.
To extract these effects, it is useful to make a coordinate transformation to a more natural metric.
There are two related reasons why the metric in (\ref{eq:metric}) is not natural.
First, the Christoffel symbols are not continuous. Second, what jumps when we cross the horizon is not the mass of the black hole,
as parameterized in (\ref{bhs}), but the coefficient that multiplies $x^+ x^-$, which is related to $Q^2$.
To fix these issues, we make the following transformation:
\be
\tilde{x}^+ = x^+ -\frac{\Theta(x^-)x^+}{2\pi\a'QM}~, ~~~~\tilde{x}^-=x^-~.\label{shockwave}\ee
This gives the following metric:
\be ds^2 \cong  \left( \frac{M}{Q}+\frac{\Theta(x^-)}{2\pi\a'Q^2}-Q^2x^-\tilde{x}^+\right)^{-1} dx^-d\tilde{x}^+ - \frac{\delta(x^-)\tilde{x}^+}{2\pi\a'M^2}(dx^-)^2~,\ee
with continuous Christoffel symbols and a jump in the mass parameter.~\footnote{The dilaton
in (\ref{eq:metric}) also becomes smooth after the transformation (\ref{shockwave}), as it should.}
Note that the mass inside the black hole is larger, since the null flux at the horizon is negative.
The last term describes a shock wave at the horizon, with 
\be
\Delta(x^+)=-\frac{x^+}{2\pi\a'QM}~,
\ee
which implies that the shift in $v\cong x^+$, when crossing $u\cong x^-=0$, due to a single folded string, is
\be \frac{\Delta v}{v} =\frac{\Delta (x^+)}{x^+} =  -\frac{1}{2\pi\a' QM} =-\frac{g_s^2}{2\pi\a'Q^2}~. \label{eq:shift} \ee
For multiple IFS's, the discontinuity is
\be \frac{\Delta v}{v} =-\frac{g_s^2}{2\pi\a'Q^2}N_{IFS}~. \ee
Hence, (\ref{clo}) gives (\ref{number}).

\end{appendix}

\end{document}